\documentclass[twocolumn,english]{revtex4}
\usepackage[T1]{fontenc}
\usepackage[latin9]{inputenc}
\usepackage{textcomp}
\usepackage{amstext}
\usepackage{amssymb}
\usepackage{graphicx}
\usepackage{babel}
\usepackage{ulem}

\usepackage[colorlinks,linkcolor=blue, urlcolor=blue, anchorcolor=blue, citecolor=blue]{hyperref}

\makeatletter

\newcommand{\lyxmathsym}[1]{\ifmmode\begingroup\def\b@ld{bold}
  \text{\ifx\math@version\b@ld\bfseries\fi#1}\endgroup\else#1\fi}

\providecommand{\tabularnewline}{\\}

\@ifundefined{textcolor}{}
{%
 \definecolor{BLACK}{gray}{0}
 \definecolor{WHITE}{gray}{1}
 \definecolor{RED}{rgb}{1,0,0}
 \definecolor{GREEN}{rgb}{0,1,0}
 \definecolor{BLUE}{rgb}{0,0,1}
 \definecolor{CYAN}{cmyk}{1,0,0,0}
 \definecolor{MAGENTA}{cmyk}{0,1,0,0}
 \definecolor{YELLOW}{cmyk}{0,0,1,0}
 }

\makeatother

\begin{document}

\title{Ba(Zn,Co)$_{2}$As$_{2}$: a II-II-V Diluted Ferromagnetic Semiconductor with N-type Carriers}

\author{Shengli Guo$^{1}$, Huiyuan Man$^{1}$, Cui Ding$^{1}$, Yao Zhao$^{1}$, Licheng Fu$^{1}$, Yilun Gu$^{1}$, Guoxiang Zhi$^{1}$, Benjamin A. Frandsen$^{2}$, Sky C. Cheung$^{2}$, Zurab Guguchia$^{2}$, Kohtaro Yamakawa$^{2}$, Bin Chen$^{3}$, Hangdong Wang$^{3}$, Z. Deng$^{4}$, C.Q. Jin$^{4}$, Yasutomo J. Uemura$^{2}$ and Fanlong Ning$^{1,5}$}

\email{ningfl@zju.edu.cn}

\selectlanguage{english}%

\affiliation{$^{1}$Department of Physics, Zhejiang University, Hangzhou 310027, China}
\affiliation{$^{2}$Department of Physics, Columbia University, New York, New York 10027, USA}
\affiliation{$^{3}$Department of Physics, Hangzhou Normal University, Hangzhou 310016, China}
\affiliation{$^{4}$Beijing National Laboratory for Condensed Matter Physics, and Institute of Physics, Chinese Academy of Sciences, Beijing 100190, China}
\affiliation{$^{5}$Collaborative Innovation Center of Advanced Microstructures, Nanjing 210093, China}

\begin{abstract}
Diluted ferromagnetic semiconductors (DMSs) that combine the properties of semiconductors with ferromagnetism have potential application in spin-sensitive electronics (spintronics) devices. The search for DMS materials exploded after the observation of ferromagnetic ordering in III-V (Ga,Mn)As films. Recently, a series of DMS compounds isostructural to iron-based superconductors have been reported. Among them, the highest Curie temperature $T_C$ of 230 K has been achieved in (Ba,K)(Zn,Mn)$_2$As$_2$. However, most DMSs, including (Ga,Mn)As, are p-type, i.e., the carriers that mediate ferromagnetism are holes. For practical applications, DMS with n-type carriers are also advantageous. Here we report the successful synthesis of a II-II-V diluted ferromagnetic semiconductor with n-type carriers, Ba(Zn,Co)$_{2}$As$_{2}$. Magnetization measurements show that the ferromagnetic transition occurs up to $T_{C} \sim$ 45 K. Hall effect and Seebeck effect measurements jointly confirm that the dominant carriers are electrons. Through muon spin relaxation ($\mu$SR), a volume sensitive magnetic probe, we have also confirmed that the ferromagnetism in Ba(Zn,Co)$_{2}$As$_{2}$ is intrinsic and the internal field is static.

\end{abstract}


\maketitle

The combination of spin and charge degrees of freedom in diluted magnetic semiconductors (DMSs) makes them promising materials in spintronics. The observation of ferromagnetism in Mn doped III-V GaAs has therefore attracted extensive attention in last two decades \cite{GaMnAs1,GaMnAs,RMPDO3,RMPDO2}. (Ga,Mn)As films are typically fabricated via low-temperature molecular beam epitaxy (LT-MBE), where Mn$^{2+}$ substitution for Ga$^{3+}$ introduces both spins and holes simultaneously. Despite the controversy about the origin of ferromagnetism in (Ga,Mn)As\cite{RMPDO1}, it has been widely accepted that the itinerant carriers mediate the ferromagnetic interaction between spatially separated magnetic ions. To date, the Curie temperature $T_{C}$ in (Ga,Mn)As has reached a maximum of $\sim$ 190-200 K \cite{190K,191K,200K}, which is still far below room temperature and therefore limits the possibilities for practical applications. Recently, a series of DMS materials that are structural derivatives of iron-based superconductors have been synthesized, including I-II-V Li(Zn,Mn)As \cite{Li(ZnMn)As}, ``1111'' (La,Ba)(Zn,Mn)AsO\cite{(LaBa)(ZnMn)AsO} and II-II-V (Ba,K)(Zn,Mn)$_{2}$As$_{2}$\cite{(BaK)(ZnMn)2As2}. Of these, (Ba,K)(Zn,Mn)$_{2}$As$_{2}$ has the highest $T_{C}$ of $\sim$ 230 K\cite{230K}, exceeding the highest record of (Ga,Mn)As. (Ba,K)(Zn,Mn)$_{2}$As$_{2}$ was synthesized through the doping of K and Mn into the parent semiconductor BaZn$_{2}$As$_{2}$, where the substitution of Mn for Zn introduces magnetic moments and the substitution of K for Ba introduces carriers. Considering that the end member BaMn$_{2}$As$_{2}$ is an antiferromagnet with Neel temperature of 625 K \cite{BaMn2As2}, it seems promising that $T_{C}$ may reach room temperature in II-II-V systems when the synthesis conditions and the selection of elements are optimzed\cite{IEEE}.

The above-mentioned DMSs are all p-type, i.e., the dominant carriers are holes. N-type DMSs with electron carriers are still exceptionally rare. In practical applications, both p- and n-type DMSs are required to fabricate junctions and devices. Furthermore, n-type DMSs may shed light on the general mechanism for ferromagnetic ordering in DMSs. In the past, Co:ZnO films have been proposed to be a candidate for n-type DMS \cite{ZnO1,ZnO2,ZnO3}. However, the underlying mechanism is still in debate. For example, one careful investigation showed that the ferromagnetism may arise from a hydrogen-facilitated interaction\cite{ZnO4}. Co:TiO$_{2}$ films are also reported to possess ferromagnetism above room temperature, with electrons provided by defects or electric fields acting as carriers\cite{TiO2_defects,TiO2_electricity}. Recently, Hai $et. \ al.$ reported the observation of electron-mediated ferromagnetism in (In,Fe)As films where interstitial Be provides electrons\cite{(InFe)As_growth,(InFe)As_magnetoresistance,(InFe)As_electronmass}. Similar fabrication routes have also been tried in (In,Co)As films, but no ferromagnetic ordering has been observed\cite{(InCo)As}. Theoretically, Gu $et. \ al.$ predicted that n-type DMSs may be realized in narrow-band-gap semiconductors\cite{Gu}.

\begin{center}
\begin{figure*}
\includegraphics[width=15cm]{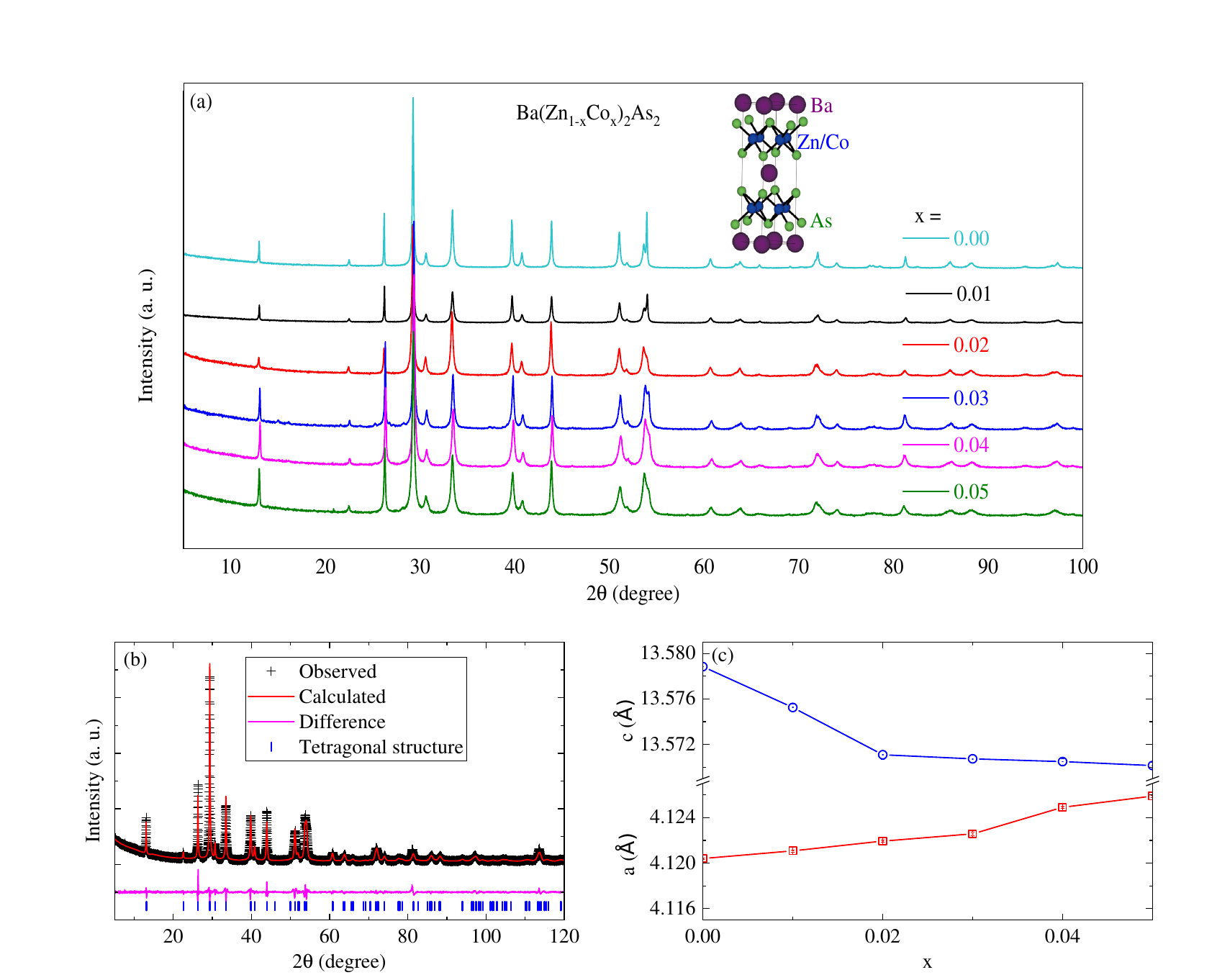}

\caption{\textbf{Structural and X-ray diffraction results.} (a) X-ray diffraction patterns of Ba(Zn$_{1-x}$Co$_{x}$)$_{2}$As$_{2}$ with different doping levels. (b) Rietveld refinement profile for $x = 0.04$. (c) Lattice parameters for Ba(Zn$_{1-x}$Co$_{x}$)$_{2}$As$_{2}$.}

\end{figure*}
\end{center}

In this paper, we demonstrate the successful synthesis of a high-quality n-type ferromagnetic semiconductor by doping Co onto the Zn sites of the narrow-band-gap (0.2 eV) semiconductor BaZn$_{2}$As$_{2}$ \cite{BaZn2As2_gap}. The highest $T_{C}$ of Ba(Zn$_{1-x}$Co$_x$)$_{2}$As$_{2}$ reaches $\sim$ 45 K for $x=0.04$. Using muon spin relaxation ($\mu$SR) measurements, we have confirmed the homogeneous and intrinsic nature of the ferromagnetic ordering in Ba(Zn,Co)$_{2}$As$_{2}$.

\section*{Results}

\begin{center}
\begin{figure*}
\includegraphics[width=15cm]{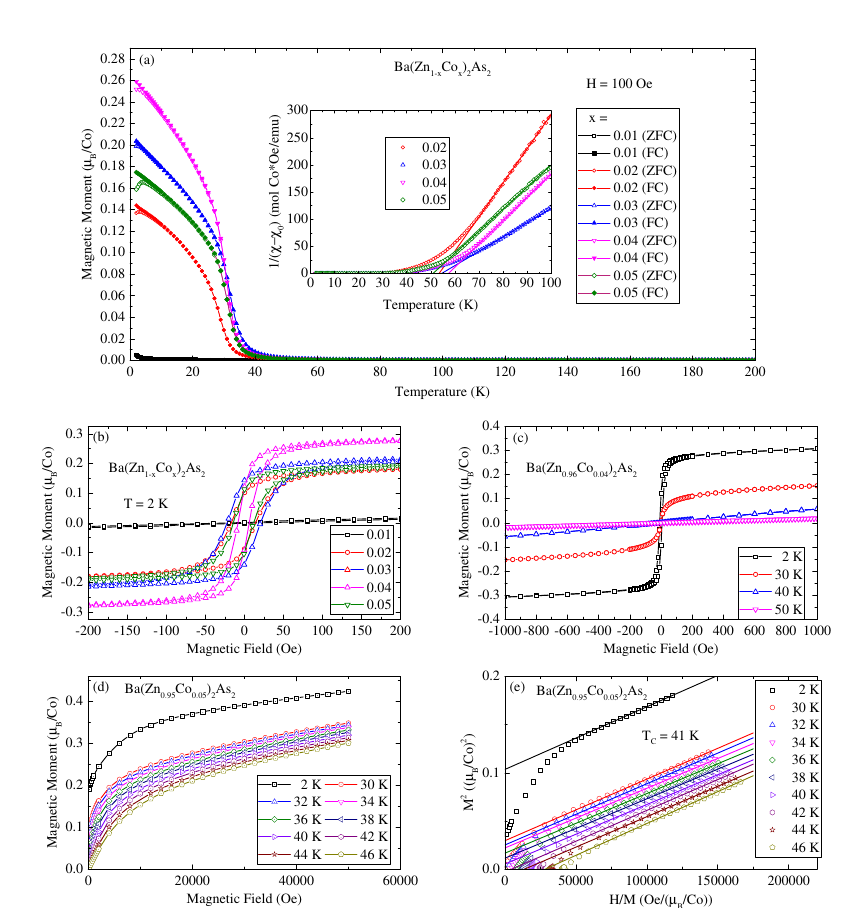}

\caption{\textbf{Magnetization results.} (a) The temperature-dependent magnetization for Ba(Zn$_{1-x}$Co$_{x}$)$_{2}$As$_{2}$ ($x=0.01,0.02,0.03,0.04,0.05$) in a magnetic field of 100 Oe. The open and filled symbols represent the zero-field-cooled and field-cooled data, respectively. Inset: Plot of $1/(\chi-\chi_{0})$ versus $T$. Straight lines represent a Curie-Weiss fit. (b) The isothermal magnetization for Ba(Zn$_{1-x}$Co$_{x}$)$_{2}$As$_{2}$ ($x=0.01,0.02,0.03,0.04,0.05$) at 2 K in an applied magnetic field ranging from -200 Oe to 200 Oe. (c) Evolution of the hysteresis loop for Ba(Zn$_{0.96}$Co$_{0.04}$)$_{2}$As$_{2}$ with increasing temperature. (d) Isothermal magnetization for Ba(Zn$_{0.95}$Co$_{0.05}$)$_{2}$As$_{2}$ at different temperatures. (e) The Arrott plot for Ba(Zn$_{0.95}$Co$_{0.05}$)$_{2}$As$_{2}$ at different temperatures. Lines show the best linear fit.}

\end{figure*}
\end{center}

\textbf{X-ray diffraction.}
In Fig. 1(a), we show the X-ray diffraction patterns for Ba(Zn$_{1-x}$Co$_{x}$)$_{2}$As$_{2}$ with different doping levels. In general, BaZn$_{2}$As$_{2}$ is polymorphic, typically crystallizing into either an orthorhombic structure (space group \textit{Pnma}) or a tetragonal structure (space group \textit{I4/mmm})\cite{(BaK)(ZnMn)2As2}. The tetragonal structure results in a semiconductor with a band gap of $\sim$ 0.2 eV\cite{BaZn2As2_gap} that forms the parent compound of the (Ba,K)(Zn,Mn)$_{2}$As$_{2}$ DMS system. In this structure, layers of \{ZnAs$_{4}$\} tetrahedra stack alternately with Ba layers along the c axis. We note that if the Zn atoms are replaced by Fe, BaFe$_{2}$As$_{2}$ is the parent compound for many iron based superconductors\cite{BaFe2As2}. The X-ray diffraction peaks in Fig. 1(a) can be well indexed with a tetragonal structure (space group \textit{I4/mmm}) with no sign of the orthorhombic phase or other impurities. In Fig. 1(b), we show the Reitveld refinement profile for the $x = 0.04$ sample using the GSAS-\uppercase\expandafter{\romannumeral2} package\cite{GSAS}. No obvious impurity peak was observed, and the resulting weighted reliability factor $R_{wp}$ is $\sim 9.87\%$, indicating high sample quality. In Fig. 1(c), we show the lattice parameters for different doping levels. With increasing Co concentration $a$ increases and $c$ decreases monotonically. The monotonic behavior of lattice parameters indicates the successful doping of Co up to $x = 0.05$.

\textbf{Magnetic properties.}
In Fig. 2(a), we show the temperature-dependent magnetization of Ba(Zn$_{1-x}$Co$_{x}$)$_{2}$As$_{2}$ ($x=0.01,0.02,0.03,0.04,0.05$) in an applied magnetic field of 100 Oe. Zero field cooling (ZFC) and field cooling (FC) data are represented by open and filled symbols, respectively. For $x=0.01$, no magnetic transition was observed down to the base temperature of 2 K, and the magnetic moment at 2 K in $H=100$ Oe is only 0.005 $\mu$B/Co. However, for Co concentrations exceeding $1\%$, a sudden increase of the magnetization develops around 35-45 K, indicative of a ferromagnetic transition.

We used the Arrott plot method for precise determination of the Curie temperature $T_{C}$ for Ba(Zn$_{1-x}$Co$_{x}$)$_{2}$As$_{2}$ \cite{Arrott}. According to the Ginzburg-Landau mean field theory for magnetism, the free energy close to the phase transition can be written as $F(M)=-HM+a(T-T_{C})M^2+bM^4+\cdots$. For a stable state, the derivative of $F(M)$ with respect to $M$ should be $0$. After omitting the high order items, the function is rewritten as: $M^{2}=\frac{1}{2b}\frac{H}{M}-\frac{a}{2b}(T-T_{C})$. Therefore, around $T_{C}$, the plot of $M^{2}$ versus $H/M$ should be an array of parallel lines. The intercept is positive below $T_{C}$ and negative above $T_{C}$, respectively. The temperature where the line passes through the origin is $T_{C}$. In Fig. 2(e), we show the Arrott plot for $x=0.05$. Around $T_{C}$, the points at high magnetic field fall approximately on a series of parallel lines. The solid lines displayed on the plot are the linear fits at high magnetic field, and the nonlinear behavior at low field is ascribed to the higher-order terms we omitted in the analysis or other deviations from mean field theory. We identify $T_C$ as 41 K, the temperature at which the parallel line would pass through the origin. $T_{C}$ for other doping levels was also determined by this method (see Supplement). We list $T_{C}$ for Ba(Zn$_{1-x}$Co$_{x}$)$_{2}$As$_{2}$ in Table 1. We can also obtain the effective moments $\mu_{eff}$ by fitting the temperature dependent magnetization above $T_{C}$ with a modified Curie-Weiss law: $\chi=\chi_{0}+C/(T-\theta)$, where $\chi_{0}$ is the temperature independent component, $C$ is the Curie constant and $\theta$ is the Weiss temperature. $\mu_{eff}$ is $\sim$ 1.1-1.7 $\mu_{B}$/Co. According to $\mu_{eff}=\mu_{B}g\sqrt{S(S+1)}$, where $\mu_{B}$ is the Bohr magneton and $g=2$ is the Lande factor for electrons, we estimate the average spin state of Co to be close to S = $1/2$.

\begin{table}
\caption{Curie Temperature ($T_{C}$), Weiss Temperature
 and Effective Moment ($\theta$ and $\mu_{eff}$, derived from Curie-Weiss fitting), Saturation Moment ($\mu_{s}$, the value measured at $T = 2$ K and $H = 200$ Oe), Coercive Field ($H_{c}$)}
\begin{tabular}{c  c c c c c}

\hline
Co concentration ($x$)& $T_{C}$ & $\theta$ & $\mu_{eff}$ & $\mu_{s}$ & $H_{c}$
\tabularnewline
& (K) & (K) & ($\mu_{B}$/Co)  & ($\mu_{B}$/Co) & (Oe)\tabularnewline
\hline 0.01 & $\slash$ & 0.03 & 2.0 & $\slash$ & $\slash$\tabularnewline
\hline 0.02 & 35 & 53 & 1.1 & 0.18 & 16\tabularnewline
\hline 0.03 & 37 & 54 & 1.7 & 0.20 & 22\tabularnewline
\hline 0.04 & 45 & 57 & 1.4 & 0.24 & 6 \tabularnewline
\hline 0.05 & 41 & 51 & 1.4 & 0.22 & 11 \tabularnewline
\hline
\end{tabular}

\caption{Comparison of selected properties of (Ga,Mn)As, (Ba,K)(Zn,Mn)$_{2}$As$_{2}$ and Ba(Zn,Co)$_{2}$As$_{2}$.}
\begin{tabular}{c c c c}
\hline
 & \tiny{(Ga,Mn)As} & \tiny{(Ba,K)(Zn,Mn)$_{2}$As$_{2}$} & \tiny{Ba(Zn,Co)$_{2}$As$_{2}$}
\tabularnewline
\hline \tiny{Valence before doping} & \uppercase\expandafter{\romannumeral3}-\uppercase\expandafter{\romannumeral5} & \uppercase\expandafter{\romannumeral2}-\uppercase\expandafter{\romannumeral2}-\uppercase\expandafter{\romannumeral5} &\uppercase\expandafter{\romannumeral2}-\uppercase\expandafter{\romannumeral2}-\uppercase\expandafter{\romannumeral5}\tabularnewline
\hline \tiny{Carrier type} & holes & holes & electrons \tabularnewline
\hline \tiny{Maximum} $T_{C}$ & 190 K\cite{190K} & 230 K\cite{230K} & 45 K \\
\hline \tiny{Saturation moment} & 5 $\mu_{B}$/Mn & 2 $\mu_{B}$/Mn & 0.2 $\mu_{B}$/Co\\
\hline \tiny{Sample form} & thin film & bulk form & bulk form\\
\hline
\end{tabular}

\end{table}

In Fig. 2(b), we show the isothermal magnetization at 2 K. Clear hysteresis loops are observed for all doping levels except the paramagnetic $x=0.01$ sample. The coercive field of Ba(Zn,Co)$_{2}$As$_{2}$ is on the order of $\sim$ 10 Oe, which is much smaller than 1 T in (Ba,K)(Zn,Mn)$_{2}$As$_{2}$\cite{(BaK)(ZnMn)2As2}. The small coercive field is consistent with the minimal bifurcation of ZFC and FC curves at 100 Oe shown in Fig. 1(a). In Fig. 2(c), we show the temperature dependence of the hysteresis loop for $x=0.04$. With increasing the temperature, the moment become smaller and the hysteresis loop eventually disappears above 50 K. The saturation moment ($\mu_{s}$) is $\sim$ 0.2 - 0.3 $\mu_{B}$/Co for Ba(Zn,Co)$_{2}$As$_{2}$ which is much smaller than $2\ \mu_{B}$/Mn for (Ba,K)(Zn,Mn)$_{2}$As$_{2}$ and $5\ \mu_{B}$/Mn  for (Ga,Mn)As\cite{GaMnAs,(BaK)(ZnMn)2As2}.

\begin{center}
\begin{figure}
\includegraphics[width=9.5cm]{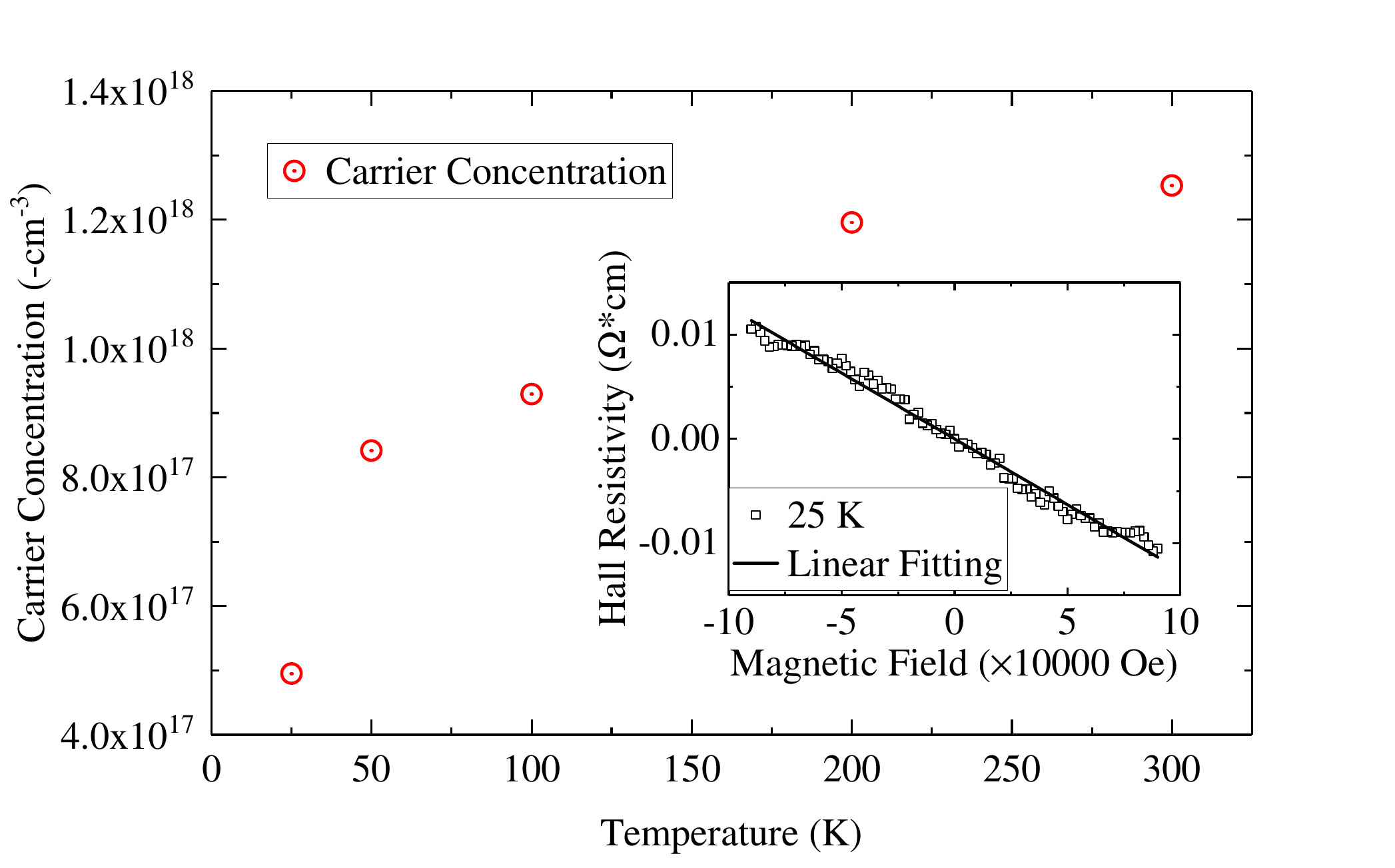}

\caption{\textbf{Results of Hall effect measurements.} Carrier concentration for Ba(Zn$_{0.96}$Co$_{0.04}$)$_{2}$As$_{2}$ calculated from Hall resistivity curves. Inset is the Hall resistivity at 25 K, with a linear fit shown as the green line.}

\end{figure}
\end{center}

\textbf{Hall effect, Seebeck effect and transport.}
We jointly utilized measurements of the Hall effect and Seebeck effect (see Supplement) to investigate the properties of the carriers. Since $R_{Hall}=B/(ne)$, where $B$ is the external field perpendicular to the current and $e$ is the elementary charge, we obtained the carrier concentration from $R_{Hall}$ versus $B$ curves. In Fig. 3, we show the representative Hall resistivity ($R_{H}$) at 25 K and the variation of carrier density ($n$) versus temperature ($T$). The negative slope of Hall resistivity curve indicates that dominant carriers in Ba(Zn,Co)$_{2}$As$_{2}$ are electrons. The carrier concentration is about 10$^{17}$ $\sim$ 10$^{18}$/cm$^{-3}$ depending on the measuring temperature, which is much smaller than 10$^{20}$/cm$^{-3}$ of (Ba,K)(Zn,Mn)$_{2}$As$_{2}$\cite{(BaK)(ZnMn)2As2}, but comparable to that of Li(Zn,Mn)P\cite{Li(ZnMn)P}. The carrier density decreases gradually with decreasing temperature. Seebeck effect measurements at room temperature were also conducted to investigate the carrier type. The Seebeck coefficient is $S=-\Delta U/\Delta T$, where $\Delta U$ is the voltage difference between two electrodes and $\Delta T$ is the temperature difference. The sign of the Seebeck coefficient is related to the carrier type, positive for p-type carriers and negative for n-type carriers. The room temperature Seebeck coefficient is $\sim$ -15.86 $\mu$V/K for $x=0.04$ and $\sim$ -6.95 $\mu$V/K for $x=0.05$. The negative Seebeck coefficient (see Supplement) confirms our conclusion of n-type carriers.

\begin{center}
\begin{figure}
\includegraphics[width=9.5cm]{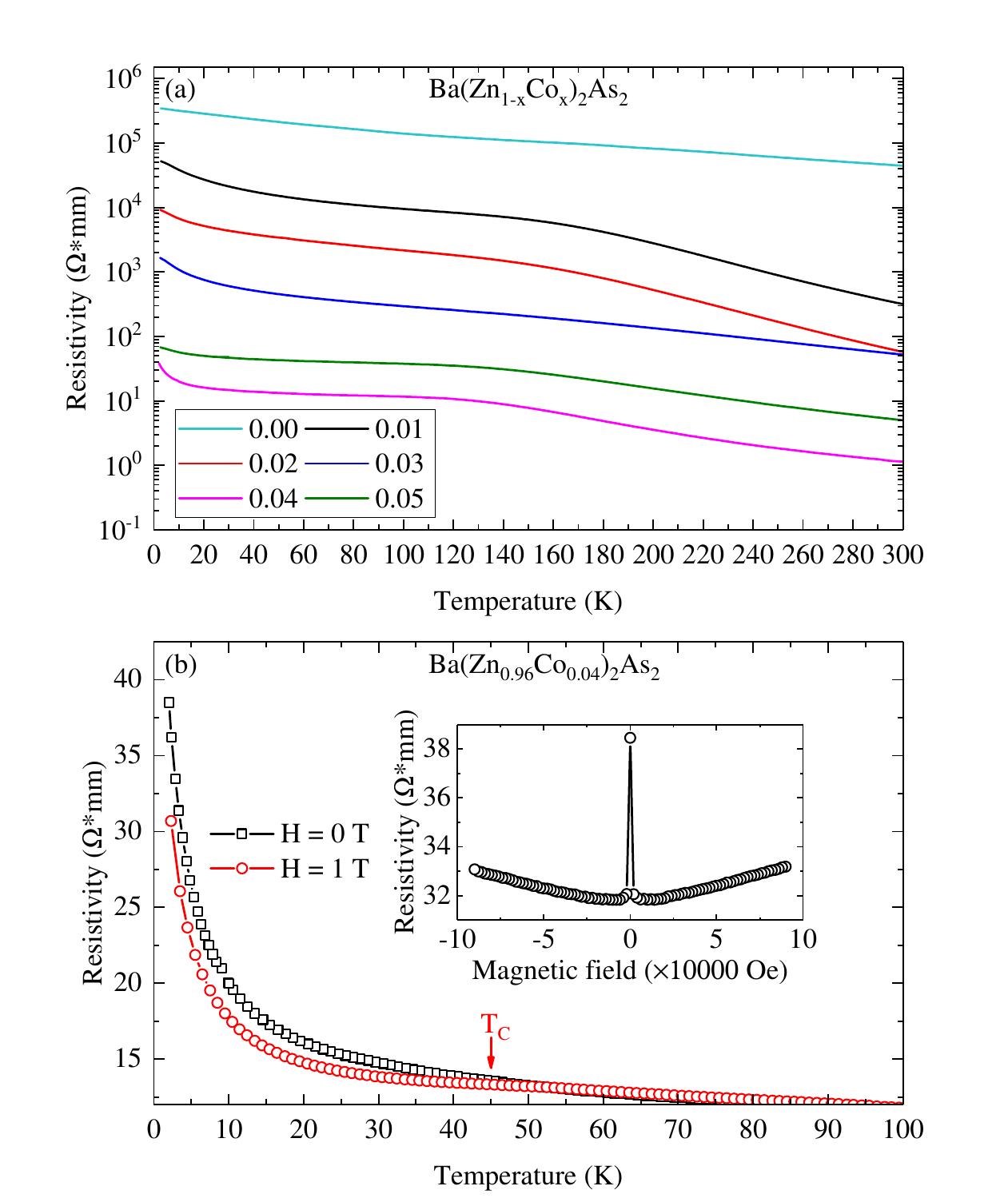}

\caption{\textbf{Results of resistivity and magneto-resistivity.} (a) The resistivity as a function of temperature for different doping levels. (b) The magneto-resistivity for the $x=0.04$ sample. The red arrow marks the position of $T_{C} \sim 45$ K. Inset is the field-dependent resistivity, with a field interval of 2000 Oe.}

\end{figure}
\end{center}

In Fig. 4(a), we show the electrical transport properties for different doping levels. With Co doping, the resistivity retains its semiconductor behavior but the magnitude decreases, indicating the successful introduction of carriers by Co substitution for Zn. In Fig. 4(b), we show the magneto-resistivity (MR) for the $x=0.04$ sample under an applied magnetic field of 1 Tesla. The MR curve decreases clearly below $T_{C}$, which is due to the suppression of magnetic scattering by the external field. At $T=2$ K, the negative MR saturates at $H = 6000$ Oe with $(\rho-\rho(0))/\rho(0)$ reaching $\sim -17\%$, as shown in the inset of Fig. 4(b). This value is much larger than the value of $\sim 7.5\%$ for (Ba,K)(Zn,Mn)$_{2}$As$_{2}$ at 7 T\cite{230K}. After saturation at 6000 Oe, the MR displays a slight increase with increasing external field. Usually, depending on the configuration of the mutually perpendicular field and current, the Lorentz force can affect the electrons' path and therefore increase the resistance. Nonetheless, the relatively large negative magneto-resistivity and smaller saturation field indicate that the electrical transport properties in Ba(Zn,Co)$_{2}$As$_{2}$ can be be easily controlled by external magnetic field.

\begin{center}
\begin{figure}
\includegraphics[width=9cm]{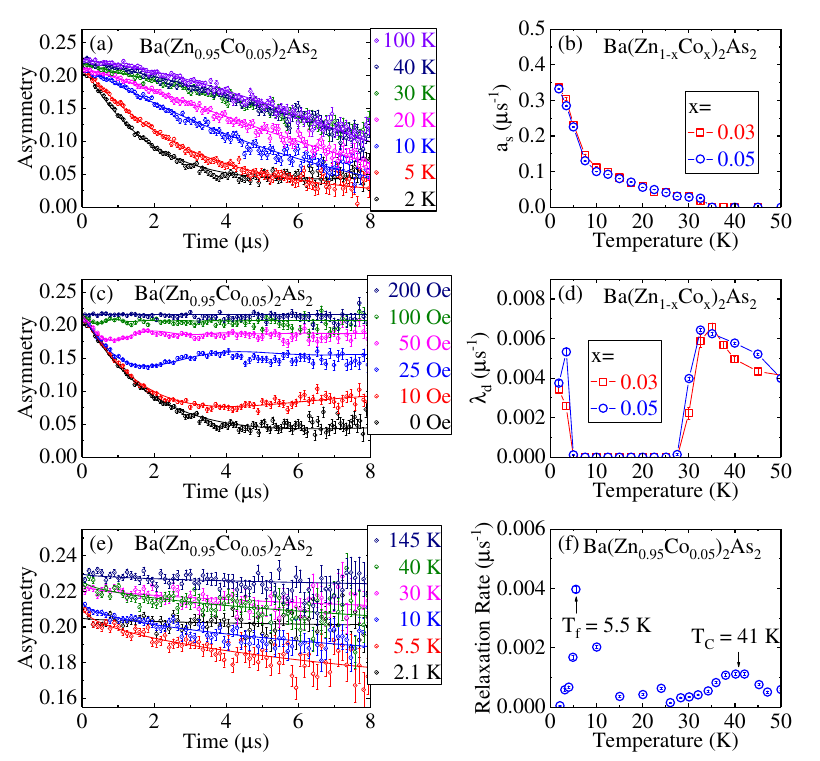}

\caption{\textbf{Results of $\mu$SR characterization.}(a) ZF-$\mu$SR time spectra of Ba(Zn$_{0.95}$Co$_{0.05}$)$_{2}$As$_{2}$. The solid lines show the best fit to the dynamic-static relaxation function with the static local field amplitude parameter $a_{s}$ shown in (b) and the dynamic relaxation rate parameter $\lambda_{d}$ shown in (d). The LF-$\mu$SR time spectra are shown in (c), exhibiting full decoupling at 200 Oe. (e) The time spectra of LF-$\mu$SR in Ba(Zn$_{0.95}$Co$_{0.05}$)$_{2}$As$_{2}$  with an external field of 100 Oe at different temperatures. (f)The muon spin relaxation rate 1/T$_1$. }

\end{figure}
\end{center}

\begin{center}
\begin{figure}
\includegraphics[width=8.5cm]{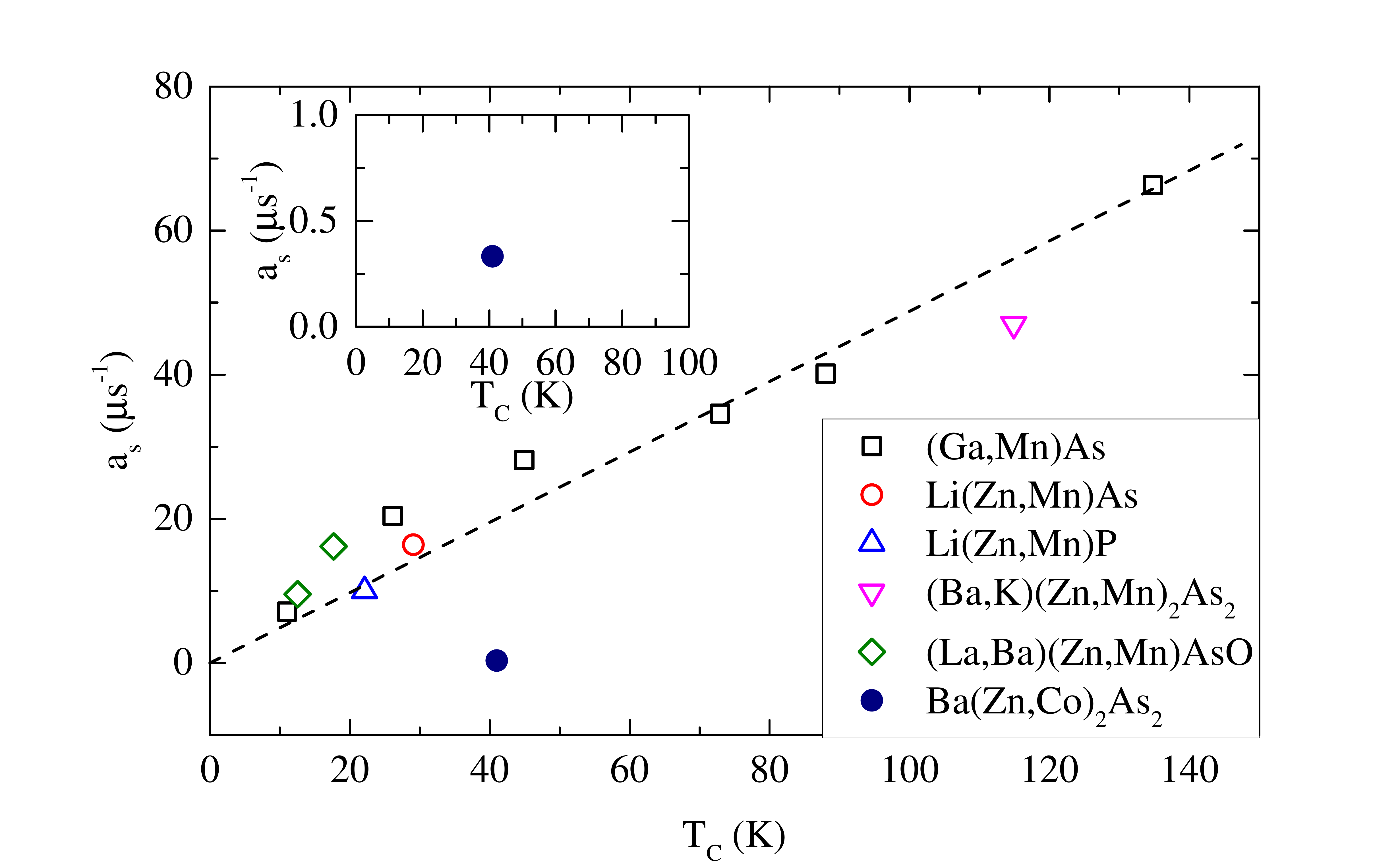}

\caption{\textbf{Plot of $a_{s}$ versus $T_{C}$.} Correlation between the static internal field
parameter $a_{s}$ determined at $T = 2$ K by ZF-$\mu$SR versus the ferromagnetic Curie temperature $T_{C}$ observed in (Ga,Mn)As (Ref. 32), Li(Zn,Mn)As (Ref. 9), Li(Zn,Mn)P (Ref. 33), (La,Ba)(Zn,Mn)AsO (Ref. 10), (Ba,K)(Zn,Mn)$_{2}$As$_{2}$ (Ref. 11), and Ba(Zn,Co)$_{2}$As$_{2}$ (current study). }

\end{figure}
\end{center}

\textbf{ZF- and LF-$\mu$SR.}
Generally speaking, a small amount of magnetic impurities such as Co nanoparticles or unknown Co compounds can give rise to magnetic signals that may obscure the intrinsic magnetic properties. To rule out such scenario, we applied $\mu$SR, a volume-sensitive magnetic probe, to investigate Ba(Zn,Co)$_{2}$As$_{2}$. In Fig. 5(a), we show the zero field (ZF-) $\mu$SR time spectra for Ba(Zn$_{0.95}$Co$_{0.05}$)$_{2}$As$_{2}$. A fast-relaxing component clearly arises below $T_{C}$, consistent with the formation of ferromagnetic ordering. Similar to the case of p-type ``1111'' DMS systems (La,Ba)(Zn,Mn)AsO\cite{(LaBa)(ZnMn)AsO}, we use a dynamic spin freezing model to fit the ZF-$\mu$SR data. As shown by the solid curves in Fig. 5(a), the time spectra can be well fitted by the dynamic-static relaxation function (Eq. (26) of Ref. 31). This indicates that Ba(Zn$_{0.95}$Co$_{0.05}$)$_{2}$As$_{2}$ achieves static magnetic order throughout the entire volume at low temperatures, confirming that the previous magnetization measurements are intrinsic to the samples and not due to a small impurity phase. The static local field amplitude $a_{s}$ and the dynamic relaxation rate $\lambda_{d}$ determined from the fits are displayed in Fig. 5(b) and (d). The parameter $a_{s}$ is proportional to the individual ordered moment size multiplied by the moment concentration. $a_{s}$ is zero above $T_{C}$, and starts to increase below $T_{C}$, indicating the emergence of a static field in the ferromagnetic state.

We used longitudinal field (LF-)$\mu$SR to investigate the spin dynamics in Ba(Zn,Co)$_{2}$As$_{2}$. In Fig. 5(c), we show the field dependence LF-$\mu$SR spectra measured at 2 K. An external field of $\sim$  100 Oe fully decouples the LF-$\mu$SR time spectra, indicating that the internal magnetic field at the muon stopping sites is fully static and has a magnitude about 10 times less than the decoupling field, i.e. $\sim$ 10 Oe. In Fig.5(e), we show the temperature-dependent LF-$\mu$SR spectra conducted under a constant external field of 100 Oe and plot the extracted relaxation rate 1/T$_1$ in Fig. 5(f). 1/T$_1$ displays similar behavior as $\lambda_{d}$ obtained from ZF-$\mu$SR(Fig. 5(d)). The dynamic relaxation exhibits two peaks, one corresponding to $T_{C}$ arising from the critical slowing down of spin fluctuations near $T_{C}$, and the other arising at the temperature where ZFC and FC curves start to bifurcate as shown in Fig. 2(a). This temperature should be related to the freezing of magnetic domains.

When we plot the internal field strength $a_{s}$ versus $T_{C}$ in Fig. 6, the point
for the present n-type system lies at a location very different from the
linear trend shown by many other p-type DMS systems \cite{Li(ZnMn)As,(LaBa)(ZnMn)AsO,(BaK)(ZnMn)2As2,(GaMn)As_MuSR,Li(ZnMn)P_MuSR}.  Since the
static internal field parameter $a_{s}$ is proportional to the concentration multiplied
by the average static moment size in dilute spin systems, the trend for the
n-type system implies that $T_{C}$ is relatively high for a given size and
density of the static ordered moments.  Hence the ferromagnetic exchange
coupling is much larger in the n-type system compared to the p-type systems. (A preliminary estimation of the s-d exchange interaction in Ba(Zn$_{1-x}$Co$_{x}$)$_{2}$As$_{2}$ is much larger than 1.2 eV of (Ga,Mn)As\cite{(GaMn)As_exchange}.)
This tendency can be partly ascribed to the difference between the present Co-doped system and Mn-doped p-type 122 DMS
systems, which involve frustration because the nearest-neighbor Mn pairs are
coupled antiferromagnetically, as can be seen in BaMn$_{2}$As$_{2}$ being a strong
antiferromagnet with $T_{N} \sim 625$ K\cite{BaMn2As2}.
In contrast, BaCo$_{2}$As$_{2}$ is a paramagnet showing a tendency towards
ferromagnetic correlation\cite{BaCo2As2_NMR,BaCo2As2}.  Therefore, there is no frustration
between neighboring Co spins in the Co-doped 122 system.  This could lead to the
smaller coercive field and stronger ferromagnetic coupling in the n-type
system compared to the p-type Mn doped DMS system.  This feature may be
helpful in obtaining higher $T_{C}$ in n-type DMS systems.

\section*{Discussion}

We have successfully synthesized the ferromagnetic semiconductor Ba(Zn,Co)$_{2}$As$_{2}$ through the solid state reaction method. Hall resistivity and Seebeck coefficient measurements jointly confirmed that the carriers in  Ba(Zn$_{1-x}$Co$_{x}$)$_{2}$As$_{2}$ are electrons. Magnetization measurements showed that the highest $T_{C}$ is $\sim 45$ K for the $4\%$ doping level and the coercive field is on the order of 10 Oe. ZF- and LF-$\mu$SR measurements show that a static field arises throughout the full sample volume below $T_{C}$, with a magnitude of about 10 Oe at the muon stopping sites. In the temperature-dependent LF-$\mu$SR measurements, we observed critical slowing down of spin fluctuations around $T_{C}$ and the freezing of magnetic domains at lower temperature. Combining the ZF- and LF-$\mu$SR time spectra, we can conclude that the present n-type DMS system exhibits characteristic signatures of dynamic slowing down followed by static magnetic order, with a magnetically ordered state in the entire volume.

The n-type DMS Ba(Zn,Co)$_{2}$As$_{2}$ ($T_{C}$ = 45 K) from the current study joins several related compounds including the p-type DMS (Ba,K)(Zn,Mn)$_{2}$As$_{2}$ ($T_{C}$ = 230 K)\cite{230K}, the Fe-based superconductor Ba(Fe,Co)$_{2}$As$_{2}$ ($T_{c}$ = 25 K)\cite{BaFe2As2}, the antiferromagnetic insulator BaMn$_{2}$As$_{2}$ ($T_{N}$ = 625 K)\cite{BaMn2As2}, and the paramagnetic metal BaCo$_{2}$As$_{2}$\cite{BaCo2As2}. They all share a common tetragonal crystal structure with a lattice mismatch of less than $5\%$. Superconducting films of Ba(Fe$_{1-x}$Co$_x$)$_2$As$_2$ have been fabricated successfully with pulsed laser deposition methods by many groups in past \cite{Hosono-films}. Recently, Xiao \textit{et al} have successfully grown high-quality epitaxial films of the tetragonal $\beta$-BaZn$_2$As$_2$\cite{BaZn2As2_gap}, and Cao \textit{et al} are working on the growth of Ba(Zn,Co)$_{2}$As$_{2}$ films \cite{Caolixin}. With the progress of thin film growth, it is conceivable that various junctions and devices can be fabricated to combine n-type DMS Ba(Zn,Co)$_{2}$As$_{2}$, p-type DMS (Ba,K)(Zn,Mn)$_{2}$As$_{2}$, and the superconductor Ba(Fe,Co)$_{2}$As$_{2}$ through the As layers.

\begin{acknowledgments}
The work at Zhejiang was supported by MOST (No. 2016YFA0300402 and No. 2014CB921203), NSF of China (11574265), NSF of Zhejiang Province (No. LR15A040001 and No. LY14A040007) and the Fundamental Research Funds for the Central Universities; at Columbia by NSF (DMR 1610633 and DMREF DMR-1436095) and thank JAEA Reimei project; at IOPCAS by NSF \& MOST through research projects. F.L. Ning acknowledge helpful discussions with B. Gu, S. Maekawa, Kaiyou Wang, Hanoh Lee, Igor Mazin, Igor Zutic and Jianhua Zhao.
\end{acknowledgments}

\section*{Method}

\textbf{Material synthesis.}
Polycrystalline samples of Ba(Zn,Co)$_{2}$As$_{2}$ were synthesized through solid state reaction of high purity elements ($\geq$ 99.9$\%$) Ba, Zn, Co and As. Mixed ingredients were placed in aluminum crucibles and sealed in evacuated silica tubes. All handling of the elements was conducted in a glove box filled with high purity Ar (the content of H$_{2}$O and O$_{2}$ is less than 0.1 ppm) except the sealing of the silica tubes. The mixture was heated to 900 $^{\circ}$C for 10 h, then held at 1150 $^{\circ}$C for 24 h followed by furnace cooling. Next, the products were ground, pressed into pellets, sealed in evacuated silica tubes, then heated to 1150 $^{\circ}$C and held for over 24 hours followed by fast cooling to keep the tetragonal phase.

\textbf{Property characterization.}
Powder X-ray diffraction was performed at room temperature using a PANalytical X-ray diffractometer (Model EMPYREAN) with monochromatic Cu-K$_{\alpha1}$ radiation. The DC magnetization measurements were conducted on a Quantum Design Magnetic Property Measurement System (MPMS3). The Hall effect and magneto-resistivity were measured on a Quantum Design Physical Property Measurement System (PPMS). The Seebeck coefficient was measured at room temperature using a commercial thermopower measurement apparatus. The zero field resistivity was measured by the typical four-probe method with a Keithley 6221 DC and AC current source and Keithley 2182A nanovoltmeter. $\mu$SR measurements were performed with LAMPF spectrometer on the M20 beamline at TRIUMF, Canada, and $\mu$SR data were analyzed using the musrfit package\cite{musrfit}.

\section*{Author contributions}
F.L.N. conceived and proposed the present study and organized the research project with Y.J.U.; S.L.G. grew the materials and conducted transport and magnetization measurement with F.L.N, H.Y.M., C.D., Y.Z., L.C.F., Y.L.G, G.X.Z., B.C. and H.D.W.; Z.D. and C.Q.J. measured the Seeback effect; S.L.G, F.L.N, B.A.F, S.C.C, K.Y. and Y.J.U. worked on $\mu$SR data acquisition at TRIUMF, and S.L.G, Z.G., F.L.N. and Y.J.U. analysed the $\mu$SR spectra. The main text was drafted by F.L.N. after input from S.L.G. and Y.J.U.; Supplementary Information was drafted by S.L.G. and F.L.N.; All authors subsequently contributed to revisions of the main text and Supplementary Information.

\end{document}